%% file: duet_sep_prc.tex
\documentclass[%
 reprint,
showpacs,
 amsmath,amssymb,
 aps,
 prc,
]{revtex4-1}

\usepackage{hyperref}
\usepackage{graphicx}
\usepackage{dcolumn}
\usepackage{bm}
\usepackage{multirow}

\begin{document}

\title{Measurement of $\sigma_{\mathrm{ABS}}$ and $\sigma_{\mathrm{CX}}$ of $\pi^+$ on carbon by DUET}%

\author{E. S. Pinzon Guerra$^1$, S. Bhadra$^1$, S. Berkman$^2$, C. Cao$^2$, P. de Perio$^3$, Y. Hayato$^4$, K. Ieki$^5$, M. Ikeda$^4$, Y. Kanazawa$^6$,  J. Kim$^2$, P. Kitching$^7$, K. Mahn$^8$, T. Nakaya$^5$,  M. Nicholson$^7$, K. Olchanski$^7$, S. Rettie$^{2,7}$, H. A. Tanaka$^3$, S. Tobayama$^2$, M. J. Wilking$^9$, T. Yamauchi$^5$, S. Yen$^7$, M. Yokoyama$^6$}

\affiliation{%
$^1$York University, Department of Physics and Astronomy, Toronto, Ontario, Canada\\
$^2$University of British Columbia, Department of Physics and Astronomy, Vancouver, British Columbia, Canada\\
$^3$University of Toronto, Department of Physics, Toronto, Ontario, Canada\\
$^4$University of Tokyo, Institute for Cosmic Ray Research, Kamioka Observatory, Kamioka, Japan\\
$^5$Kyoto University, Department of Physics, Kyoto, Japan\\
$^6$University of Tokyo, Department of Physics, Tokyo, Japan\\
$^7$TRIUMF, Vancouver, British Columbia, Canada\\
$^8$Michigan State University, Department of Physics and Astronomy, East Lansing, Michigan, U.S.A\\
$^9$State University of New York at Stony Brook, Department of Physics and Astronomy, Stony Brook, New York, U.S.A.
}%

\collaboration{DUET Collaboration}

\date{\today}

\begin{abstract}

The DUET Collaboration reports on the measurements of the absorption ($\sigma_{\mathrm{ABS}}$) and charge exchange ($\sigma_{\mathrm{CX}}$) cross sections of positively charged pions on carbon nuclei for the momentum range 201.6 MeV$/c$ to 295.1 MeV$/c$. The uncertainties on the absorption and charge exchange cross sections are $\sim$9.5\% and $\sim$18\%, respectively. The results are in good agreement with previous experiments. A covariance matrix correlating the 5 $\sigma_{\mathrm{ABS}}$ and 5 $\sigma_{\mathrm{CX}}$ measured data points is also reported.

\end{abstract}

\maketitle


\input{introduction.tex}
\input{experimental.tex}
\input{physics.tex}
\input{event_selection.tex}
\input{cross_section.tex}
\input{uncertainties.tex}
\input{result.tex}

\begin{acknowledgments}
We are grateful for all the technical and financial support received from TRIUMF.
E. S. Pinzon Guerra acknowledges support through the Ontario Graduate Scholarship.
This work was supported by JSPS KAKENHI Grants Number 22684008, 26247034, 18071005, 20674004
and the Global COE program in Japan. M. Ikeda and K. Ieki would like to acknowledge support from JSPS.
We acknowledge the support from the NSERC Discovery Grants program, the Canadian Foundation for Innovation’s
Leadership Opportunity Fund, the British Columbia Knowledge Development Fund, and NRC in Canada.
Computations were performed on the GPC supercomputer at the SciNet HPC Consortium \cite{scinet}
SciNet is funded by: the Canada Foundation for Innovation under the auspices of Compute Canada; 
the Government of Ontario; Ontario Research Fund - Research Excellence; and the University of Toronto.
\end{acknowledgments}


\bibliography{paperNotes}

\end{document}

%% file: introduction.tex
\section{\label{sec:intro}Introduction\protect}
The scattering of pions off of atomic nuclei has been the subject of extensive study
due to its ability to serve as a probe of the nuclear structure 
through the understanding of the interactions among mesons and nucleons. The $\Delta(1232)$ pion-nucleon resonance dominates in the sub-GeV energy region, and thus the range of $p_{\pi}$ between 200 to 300 MeV$/c$ is of special interest.

\begin{figure}[ht]
\includegraphics[width=86mm]{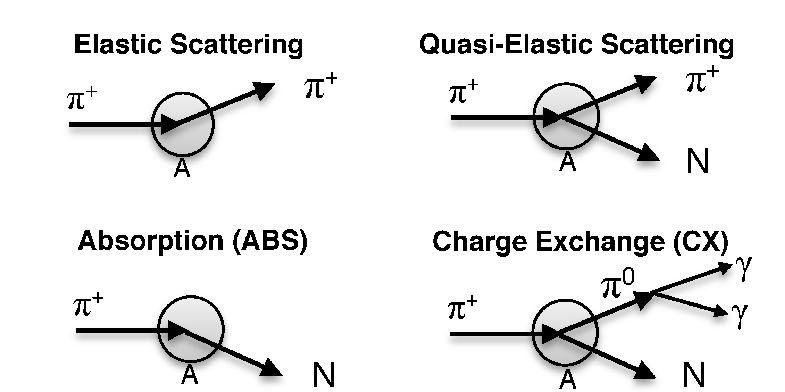}
\caption{Dominant $\pi^{\pm}$-A interactions in the sub-GeV region. ``N" represents any number of nucleons leaving the nucleus.}
\label{fig:interactions}
\end{figure}

The dominant $\pi^{\pm}$-A interactions in the sub-GeV region are represented diagrammatically in Fig. \ref{fig:interactions}. The total, elastic and quasi-elastic processes have been measured with $<10\%$ precision by various experiments \cite{Allardyce,Binon,Saunders,Gelderloos,Levenson,Ashery2,Ingram,Jones,Ashery,Bellotti1973,Bellotti1973_2}, however data are scarce for inelastic processes such as absorption (ABS: $\pi^{\pm}+A\rightarrow (A-N) + N$) and single charge exchange (CX: $\pi^{\pm} + A \rightarrow \pi^{0}+ (A-N) + N$), where ``N" represents any number of nucleons leaving the nucleus. Moreover, the majority of past experiments measured the combined rate of these two processes \cite{gianneli,navon}, and relied on other experimental results or on theoretical calculations to separate their individual contribution while ignoring possible correlations and systematic uncertainties.

Interest in pion inelastic interactions has increased in recent years due to the use of nuclear targets in GeV-scale neutrino experiments. Neutrinos are primarily detected via charged-current quasi-elastic interactions ($\nu_{\mu}+n\rightarrow \mu^{-} + p$) with target atomic nuclei. The neutrino-induced single pion production processes also contribute to the cross section in this energy range. The energy of the neutrino is fundamental to probing the oscillation phenomena and is inferred from the measured kinematics of the outgoing lepton.  If the pion is produced but not detected due to final-state interactions (FSI) within the target nucleus or secondary interactions (SI) elsewhere in the detectors, the inferred neutrino energy will be biased. 

FSI and SI are leading contributors to systematic uncertainties in neutrino oscillation and cross section experiments. Their impact is typically evaluated using predictions based on models implemented in Monte Carlo neutrino event generators such as \textsc{Neut} \cite{NEUT} and \textsc{NuWro} \cite{NuWro} for FSI, or detector simulation toolkits such as \textsc{Geant4} \cite{bertini} and \textsc{Fluka} \cite{fluka1,fluka2} for SI. Some of these generators use similar implementations of semi-classical cascade models in which the pion is propagated within the nucleus and its fate is calculated following theoretical optical models in which the pion-nucleus scattering is represented as a wave in a complex potential. The real part of the potential is responsible for elastic scattering while the imaginary part gives the contributions from inelastic channels \cite{Oset,Salcedo}. Precise tuning of the models is achieved through the empirical scaling of the theoretical microscopic interaction rates, relying entirely on the available $\pi^{\pm}$-A scattering data. Other important scenarios in which $\pi^{\pm}$-A interactions are relevant for neutrino physics are: i) the enhancement of the neutral-current $\pi^{0}$ background in neutrino oscillation appearance experiments, and, ii) pion reconstruction capabilities in water Cherenkov detectors via the explicit identification of their hadronic interactions.

An earlier paper from the DUET Collaboration \cite{duet} described our experimental setup and presented a measurement of the combined ABS and CX cross section $\sigma_{\mathrm{ABS}+\mathrm{CX}}$ in the 200 to 300 MeV$/c$ region. In this paper, we present separate measurements of $\sigma_{\mathrm{CX}}$ and $\sigma_{\mathrm{ABS}}$ for various momenta. This was achieved by extending the selection using a downstream detector to tag forward-going photons from the decay of a $\pi^0$ produced in a CX interaction. This measurement will help improve the modeling of FSI and SI and to reduce the associated systematic uncertainties on current and future neutrino oscillation and cross section experiments.

%% file: experimental.tex
\section{Experimental Apparatus}\label{sec:experiment}
\begin{figure}[ht]
\includegraphics[width=90mm]{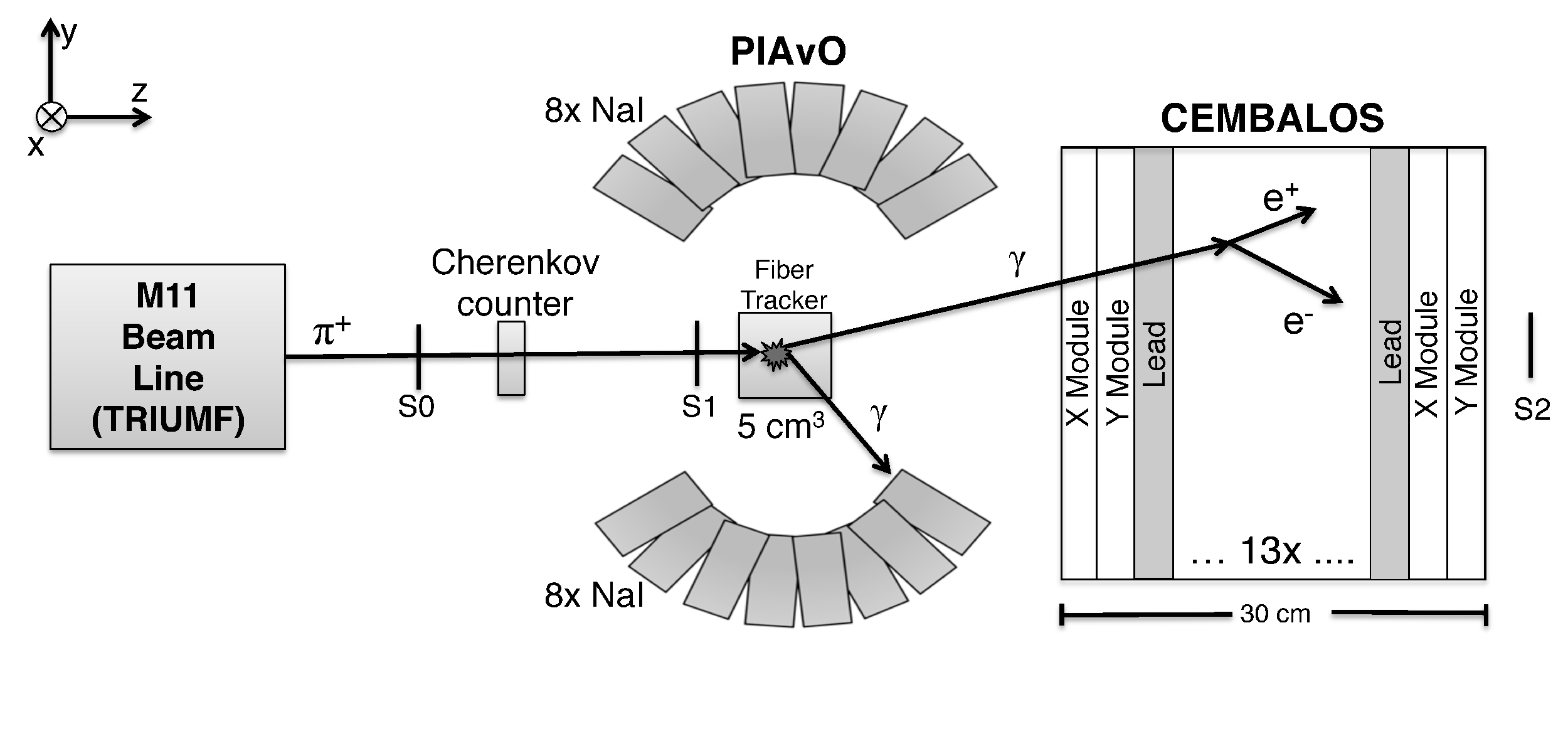}
\caption{Schematic overview of the experimental apparatus.}
\label{fig:config}
\end{figure}

The DUET experiment used the M11 beam line at TRIUMF which produced a $\pi^{+}$ beam of $>$99\% purity at five different momentum settings between 201.6 MeV/$c$ and 295.5 MeV/$c$. An extensive description of the beamline, beam particle identification, and the PIA$\nu$O detector can be found in Sec. II of \cite{duet}. Fig.  \ref{fig:config} shows a schematic overview of the experimental apparatus.

PIA$\nu$O (PIon detector for Analysis of neutrino Oscillation) served as an active carbon target for pion interactions and  provided excellent tracking capabilities and $dE/dx$ measurements of charged particles. It provided sufficient information to select ABS+CX events by requiring no observed $\pi^{+}$ in the final state. It consisted of 16 horizontal and 16 vertical layers of scintillating material, each with 32 fibers. The dimension of the region where the fibers cross each other was $49\times49\times51$ mm$^3$, providing $(1.518\pm0.007)\times10^{24}$ carbon target nuclei in its fiducial volume. The scintillation light from the fibers was read out by multi-anode photomultiplier tubes. 16 NaI crystal detectors were placed around the tracker region but are not used for this analysis.

The forward-going photons following the decay of a $\pi^0$ produced in a CX interaction were identified using the CEMBALOS (Charge Exchange Measurement By A Lead On Scintillator) detector.

\subsection{CEMBALOS}
The CEMBALOS detector was a scaled down (1/6) version of the Fine-Grained Detectors (FGDs) \cite{fgd} of T2K. It was located 25 cm downstream of PIA$\nu$O. The active portion of the detector was composed of scintillator bars made of polystyrene co-extruded with a 0.25 mm thick reflective coating of polystyrene mixed with TiO$_2$. The light yield from the far end of a bar was measured to be up to 16-18 photoelectrons (p.e.) for a minimum ionizing particle. The optical crosstalk through the TiO$_2$ coating between bars was measured to be 0.5$\pm$0.02\%. 

The scintillator bars were arranged into 15 XY modules oriented perpendicular to the beam. Each XY module contained 32 bars in the $x$ direction glued to 32 bars in the $y$ direction. Layers of 0.25 mm thick fiberglass (G10) were glued to both the upstream and downstream surfaces to provide support, and no adhesive was applied between the bars. Each module had dimensions of 32$\times$32$\times$2.02 cm$^3$. Unlike the FGDs, 0.8$\sim$1 mm thick lead layers were interspersed in between each module to enhance photon conversion. 


The scintillation light from each bar was collected by a 1 mm $\pm$ 2\% diameter wavelength shifting (WLS) double-clad Kuraray Y11 (200) S-35 J-type fiber inserted through an axial hole. The absorption and subsequent emission wavelengths for these fibers were 430 nm and 476 nm, respectively. Unlike the FGDs, due to limited availability only fibers in the last 3 XY modules had one of their ends mirrored to enhance light collection by aluminizing.

Multi-Pixel Photon Counters (MPPCs) manufactured by Hamamatsu Photonics (S10362-13-050C) were used as photosensors to measure the scintillation light. These provided excellent photon counting capability with higher quantum efficiency than photo-multipliers for the spectra of light produced by the WLS fibers. Its outer dimensions were 5$\times$6 mm$^2$, while the sensitive area was 1.3$\times$1.3 mm$^2$ containing 667 avalanche photo-diode pixels. The small size allowed for using one MPPC per bar, eliminating the possibility of crosstalk at the sensor. A custom connector was developed to achieve good optical coupling.
The XY modules were held rigidly in place inside an aluminum light-tight box. The read out electronics were mounted on the outer sides of the box to separate elements generating heat and to prevent temperature induced effects on the MPPCs. 


\subsubsection{\bf Detector simulation and calibration}\label{section:calibration}
The simulation of the CEMBALOS detector was based on that developed for the FGDs used by T2K. It made use of the \textsc{Geant4} version 9.4 patch 04~\cite{geant} simulation toolkit. Details of the geometry of the detector were simulated, including, but not limited to, the fiber structure (core, double cladding and coating), the G10 layers and the glue used to hold them to the fibers, and the measured thickness of the interspersed lead layers.

The energy deposit from charged particles traversing the scintillating bars was calculated from the pulse height ($PH$) of the digitized MPPC waveforms by the following procedure:

\begin{enumerate}
\item {\it Conversion from PH to photoelectrons:} The $PH$ measured in ADC units were translated into the number of photoelectrons $N_{pe}$ by normalizing to the average pulse height $\left\langle PH \right\rangle$ corresponding to a single-pixel avalanche. 
\begin{equation}
N_{pe} = PH/\left\langle PH \right\rangle
\end{equation}
The distribution of dark noise pulse heights was used to measure $\left\langle PH \right\rangle$ and it was found to be 48.65 ADC units. 
\item{\it Corrections for variations in overvoltage:} Temperature variations can change the overvoltage, the difference between the operating and breakdown voltages in the MPPCs, affecting the photon detection efficiency and the crosstalk and after-pulsing probabilities. Empirical corrections were applied to compensate for these effects.
\item{\it Correction for saturation of the MPPCs:} Since each MPPC has a finite number of pixels, the pulse height can get saturated. A correction based on an empirical exponential expression was applied.
\item{\it Correction for bar-to-bar variations:} Minor variations in the fiber-MPPC coupling, scintillation material, fiber mirroring, diameter of the hole, etc. can introduce a difference in the light yield for each bar. Those variations are accounted for by an additional correction ($C_{bar}$) representing the factor by which the efficiency for conversion of energy deposition in the bar to number of photons hitting the MPPCs differs from its value averaged over all the bars in CEMBALOS.
\item{\it Correction for light loss along the bar:} The light attenuation in each fiber was measured for both mirrored and unmirrored bars using cosmic rays. Fig. \ref{fig:attcurves} shows the resulting fitted distributions for the measured yield ($N_{DPE}$) of detectable photoelectrons as a function of the distance of the hit to the MPPC. The fit function is an empirical descriptor of the attenuation process.
\begin{figure}[!h]
\begin{center}
\includegraphics[width=85mm]{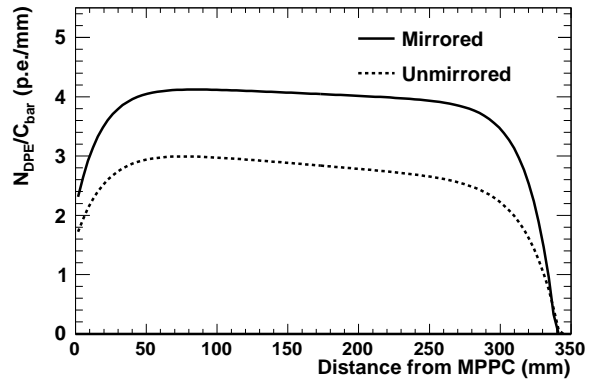}
\caption{Light attenuation curves in CEMBALOS for mirrored (solid) and unmirrored (dashed) fibers.}
\label{fig:attcurves}
\end{center} 
\end{figure}

\item{The final conversion from number of scintillation photons to energy deposition measured in p.e. involved  an empirical normalization constant and Birk's formula was used to account for the nonlinearity in the scintillator response. We adopted 0.0208$\pm$0.0003(stat)$\pm$0.0023(sys) cm/MeV for the value of Birk's constant as measured by the K2K SciBar group \cite{scibar}. A minimum of 5 p.e. was required to label an energy deposit as a hit.}
\end{enumerate}

\begin{figure}[!h]
\begin{center}
\includegraphics[width=85mm]{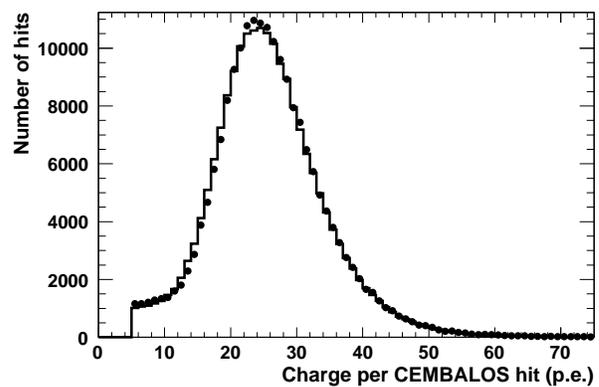}
\caption{Charge per CEMBALOS hit distribution (in photoelectrons) of through-going muons in the $p_{\pi}$=237.2 MeV$/c$ setting for data (circles) and MC (solid line), after the calibration procedure was applied. The statistical error bars are too small to appear.}
\label{fig:muoncharge}
\end{center}
\end{figure}

A control sample of beam muons in the $p_{\pi}$=237.2 MeV$/c$ setting traversing CEMBALOS was used to calibrate the charge simulation. Fig. \ref{fig:muoncharge} shows the deposited charge distribution of through-going muons for data and MC after the calibration procedure. 

\subsection{Event Summary}
The data set used in this analysis is the same as in \cite{duet}. Data were recorded from a $\pi^{+}$ beam on the PIA$\nu$O scintillator (carbon) target for five incident momenta (201.6, 216.6, 237.2, 265.5, 295.1 MeV$/c$). There were $\sim$1.5 million beam triggered events recorded for each momentum setting, except for the 216.6 MeV$/c$ setting where only 30\% was recorded due to limited beam time.

%% file: physics.tex
\section{Physics Modeling}\label{sec:physics}
\begin{figure}[h]
 \includegraphics[width=86mm]{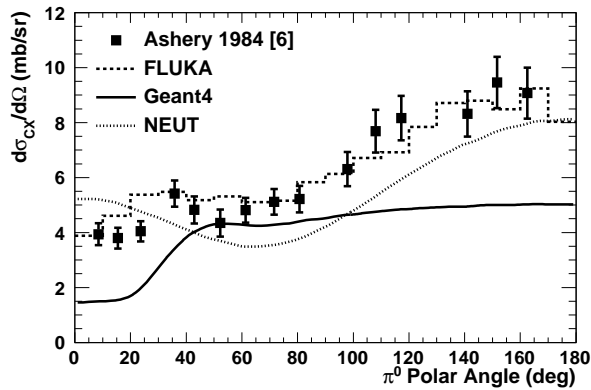}
 \caption{$d\sigma_{\mathrm{CX}}/d\Omega$ as a function of the outgoing $\pi^0$ polar angle (in the lab frame) for 265 MeV$/c$ $\pi^{+}$ interacting on $^{16}$O, for \textsc{Fluka} (dashed line), \textsc{Geant4} (solid line) and \textsc{NEUT} (dotted line), along with data from \cite{Ashery2}.}
 \label{fig:pi0kinem}
\end{figure}

The kinematics of the outgoing $\pi^0$ from CX interactions are not well known. The only existing differential cross section measurement on light nuclei from Ashery \textit{et.al.} \cite{Ashery2} is of 265 MeV$/c$ $\pi^{+}$ on oxygen. A comparison of this data and predictions from the \textsc{Neut} (v5.3.5) cascade model \cite{NEUT}, the \textsc{Geant4} (v9.04.04) Bertini cascade model \cite{bertini}, and the \textsc{Fluka} cascade model \cite{fluka1,fluka2} is shown in Fig. \ref{fig:pi0kinem}. The discrepancy among models is largest in the forward region, where CEMBALOS is most sensitive. In particular, the \textsc{Geant4} Bertini Cascade model used by our simulation shows the largest disagreement with data \cite{Ashery2}.

The modeling of the multiplicity and kinematics for nucleons ejected following an ABS or CX interaction show even larger discrepancies among models. The mechanisms for these processes are further complicated by the possibility of FSI of the nucleons before they exit the nucleus. Data on light nuclei that would help in the understanding and tuning of these processes are very scarce. \textsc{Neut} uses nucleon multiplicities published by \cite{Rowntree} of $\sigma_{\mathrm{ABS}}$ on N and Ar targets, but it is unclear what other models use.

%% file: event_selection.tex
\section{\label{sec:selection}Event selection}
The event selection described in \cite{duet} used the PIA$\nu$O detector to identify events with no $\pi^{+}$ in the final state which are consistent with ABS+CX final states. As mentioned, for the analysis presented in this paper the selection was extended by using information from the downstream detector, CEMBALOS, to identify photons following a CX interaction. 

\begin{figure}[ht]
\includegraphics[width=90mm]{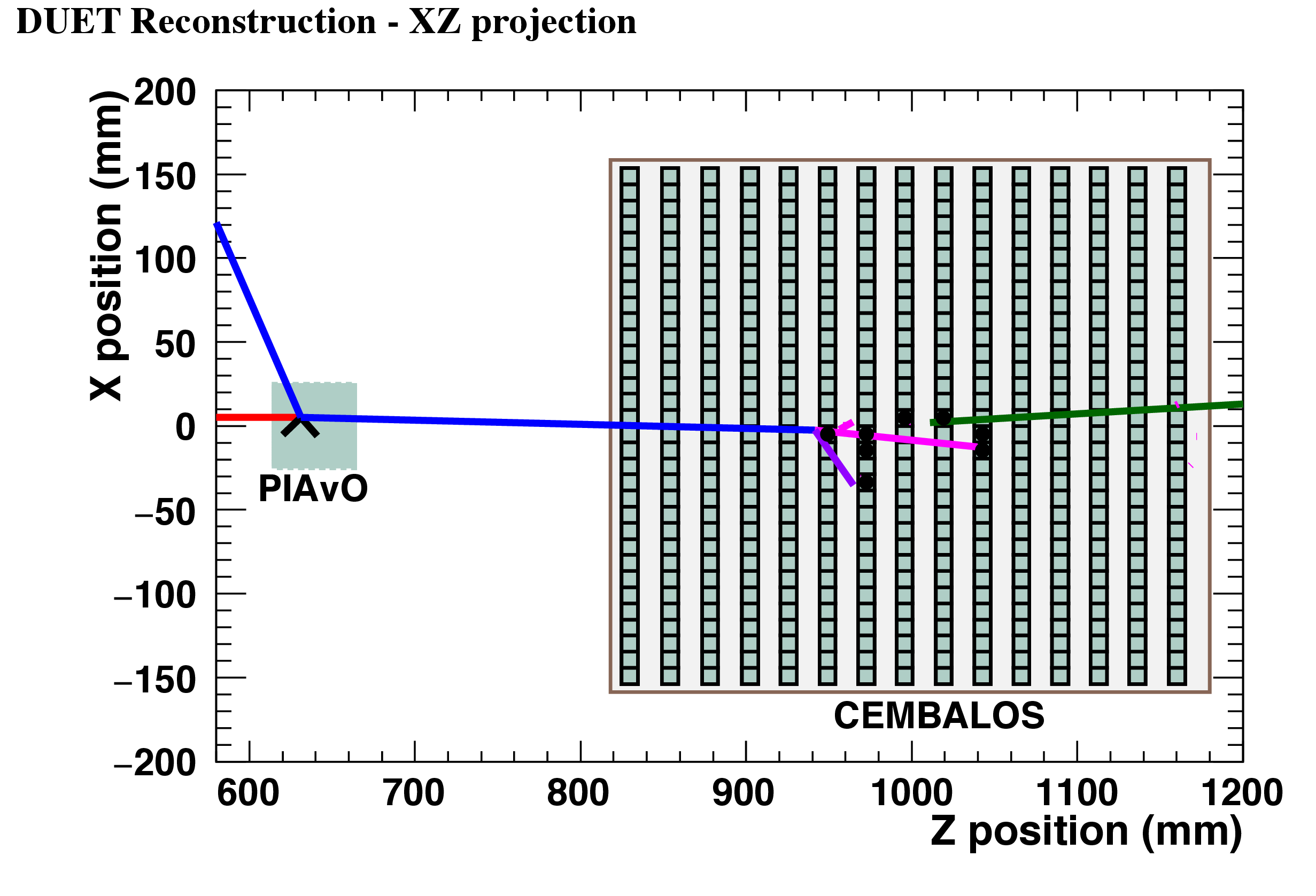}
\caption{(Color online) Example of a simulated CX event in the DUET detector setup. A 237.2 MeV$/c$ $\pi^+$ (red) undergoes CX in PIA$\nu$O producing two protons (black) and a $\pi^0$ that decay into two photons (blue). The forward-going photon is identified in CEMBALOS as it produces $e^{+}$-$e^{-}$ pairs (purple, magenta) and hits are recorded in the scintillating material.}
\label{fig:event}
\end{figure}

A simulated CX event is shown in Fig. \ref{fig:event}. The upstream horizontal (red) track represents a $\pi^{+}$ interacting in the PIA$\nu$O detector. As it undergoes a CX interaction two protons (black) and a $\pi^{0}$ are produced. The $\pi^{0}$ subsequently decays into two photons (blue). The forward-going photon travels to CEMBALOS where it converts into $e^{+}$-$e^{-}$ pairs and deposits charge in the scintillating material.

\subsection{PIA$\nu$O upstream selection}
The PIA$\nu$O detector track reconstruction algorithm used charge deposition information to reconstruct and identify charged particles in the detector and to identify an interaction vertex within a defined fiducial volume (FV). A detailed description can be found in Sec. III of \cite{duet}. A summary of the upstream selection criteria follows:
\begin{enumerate}
\item {\bf Good incident $\pi^{+}$\\}
This selection criteria is threefold: firstly, an incident $\pi^{+}$ was selected using TOF and Cherenkov light information. Secondly, a straight track, normal to the incident plane, leaving hits in the first 5 layers was required. Thirdly, this incident track was required to enter a defined fiducial volume (FV).
\item {\bf Vertex inside the FV\\}
Events with pion interactions were selected by requiring a reconstructed vertex inside the FV.  A vertex was defined as the intersection of reconstructed tracks. This removed through-going pion events, as well as small-angle pion scattering events.
\item {\bf No $\pi^{+}$ final track\\}
Reconstructed tracks exiting the interaction vertex were classified into ``proton-like" and ``pion-like" tracks using an angle-dependent cut on the deposited charge, $dQ/dx$. Events with no ``pion-like'' tracks in the final state were selected.
\end{enumerate}
The number of selected ABS+CX events, efficiency, and purity for each momentum setting can be found in \cite{duet}. About 7000 events were selected at each momentum setting, other than the shortened 216.6 MeV$/c$ momentum run, which produced only 1800 events. The efficiency and purity of the selection, $\sim$79\% and $\sim$73\%, respectively, were similar for all 5 momentum settings.

\subsection{CEMBALOS selection}
\begin{figure}[ht]
 \includegraphics[width=86mm]{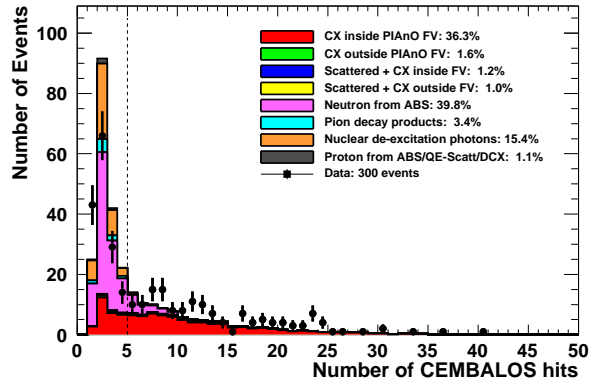}
 \caption{(Color online) Distribution of the number of hits in CEMABLOS for Data and MC (broken down into topologies and listed with their corresponding percentage composition) in the $p_\pi=$237.2 MeV$/c$ setting after applying the veto cut. Topologies contributing less than 1\% are not plotted.}
 \label{fig:nhits}
\end{figure}

\begin{figure}[ht]
 \includegraphics[width=86mm]{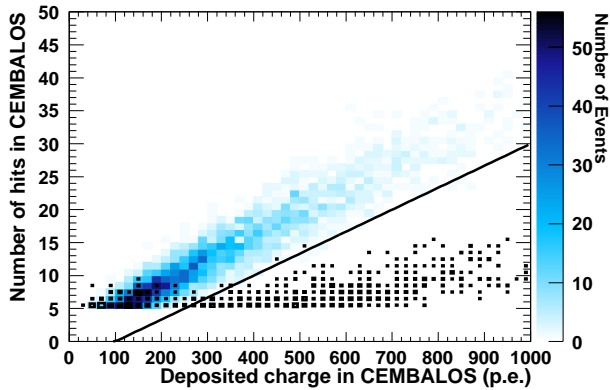}
 \caption{(Color online) Distribution of the number of hits in CEMABLOS vs. charge deposited for MC in the $p_\pi=$237.2 MeV$/c$ setting after applying the requirement of a minimum of 5 hits. The blue entries are true CX events, whereas the black boxes correspond to neutron background events.}
 \label{fig:nhitsvsCharge}
\end{figure}

Charge deposition information from CEMBALOS was used to identify CX interactions occurring in PIA$\nu$O. The main goal was to tag one of the photons from the decay of a $\pi^0$ by identifying the corresponding electromagnetic shower in CEMBALOS. The limited angular coverage ($\sim0.53 sr$) of CEMBALOS imposed the largest efficiency loss. The selection criteria were as follows:
\begin{enumerate}
\item{\bf Veto cut\\}
Charged particles in CEMBALOS left a signal in the scintillator material. Fig. \ref{fig:veto} shows the distribution of the position of the most upstream hit in CEMBALOS for each event. Each bar represents a scintillation plane. A veto cut on the first XY modules was applied to remove most of the charged particle backgrounds, such as as low-angle $\pi^+$ scatters and protons from ABS events.

\begin{figure}[ht]
 \includegraphics[width=86mm]{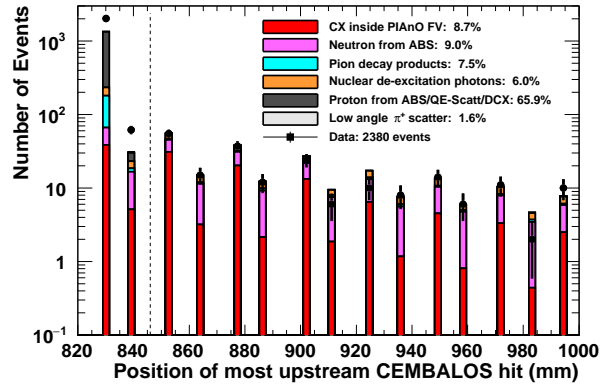}
 \caption{(Color online) Distribution of the most upstream position of CEMBALOS hits for Data and MC (broken down into topologies and listed with their corresponding percentage composition) in the $p_\pi=$237.2 MeV$/c$ setting after applying the PIA$\nu$O upstream selection. Each bar represents an XY module. Topologies contributing less than 1\% are not plotted.}
 \label{fig:veto}
\end{figure}
   
\item{\bf Hit Charge vs. Multiplicity\\}
The remaining background after the veto cut are produced by neutrons from ABS events and nuclear de-excitation $\gamma$-rays. Fig. \ref{fig:nhits} shows the distribution of the number of hits (multiplicity) in CEMBALOS. A minimum of five hits was required to reduce background from these sources. Fig. \ref{fig:nhitsvsCharge} shows the CEMBALOS hit charge vs. multiplicity distribution after applying the veto cut. A diagonal cut in this plane was applied to further reduce the remaining background of neutrons from ABS.
\end{enumerate}

\subsection{Selection purities and efficiencies}
The numbers of selected events for each momentum setting after the PIA$\nu$O and CEMBALOS selections were applied are summarized in Table \ref{tbl:short_event_summary} for data ($N_{\mathrm{Data}}$) and  \textsc{Geant4} MC (split into signal $N_{\mathrm{CX}}^{\mathrm{G4}}$ and background $N_{\mathrm{BG}}^{\mathrm{G4}})$. There are $\sim$100 events in data after the event selection, except for the 216.6 MeV$/c$ setting. The efficiencies and purities to select CX events which occurred inside the FV were around $\sim6\%$ and $\sim90\%$ respectively.  The efficiencies to select events which occurred inside the FV and had at least one of the CX photons in the direction of CEMBALOS were estimated to be $\sim30\%$.

\begin{table}[h]
   \begin{tabular}{c|ccccc}
    \noalign{\hrule height 1pt}
    $p_{\pi}$  [MeV$/c$] & $N_{\mathrm{Data}}$ & $N_{\mathrm{CX}}^{\mathrm{G4}}$ & $N_{\mathrm{BG}}^{\mathrm{G4}}$ & Efficiency [\%] & Purity [\%] \\\hline
    201.6 & 104 & 60.4 & 8.6 & 5.1 & 87.5  \\
    216.6 & 20  & 15.8 & 2.4 & 5.3 & 86.6  \\
    237.2 & 141 & 75.9 & 11.1 & 5.9 & 87.2  \\
    265.6 & 152 & 87.1 & 10.4 & 7.0 & 89.3  \\
    295.1 & 163 & 119.4 & 12.8 & 8.1 & 90.3  \\
    \noalign{\hrule height 1pt}
   \end{tabular}
\caption{Summary of number of events selected after the CEMBALOS downstream selection in Data and MC for each momentum setting, along with estimated efficiencies and purities for \textsc{Geant4}}.
\label{tbl:short_event_summary}
\end{table}
 

%% file: cross_section.tex
\section{$\sigma_{\mathrm{CX}}$ and $\sigma_{\mathrm{ABS}}$ extraction}\label{sec:xsec}
As was mentioned in Sec. \ref{section:calibration}, our simulation is based on the \textsc{Geant4} package which uses the Bertini cascade model for modeling pion inelastic interactions but also handles other complex aspects of the analysis such as the geometrical description of the detectors. In order to estimate the number of signal ($N_{\mathrm{CX}}^{\mathrm{MC}}$) and background ($N_{\mathrm{BG}}^{\mathrm{MC}}$) events predicted by the different models shown in Sec. \ref{sec:physics} without having to rewrite the simulation using each toolkit, a scheme was developed to replace the detector simulation with a set of 2D selection, rejection, and mis-reconstruction efficiencies in momentum and angle bins of the outgoing particles and presented in Sec. \ref{sec:efficiencies}. These were then applied to the predictions from \textsc{Neut} and \textsc{Fluka} obtained using thin target ($\sim$1 mm) simulations and a nominal model was selected in Sec \ref{sec:nominal}.

The measured $\sigma_{\mathrm{CX}}$ was obtained for each model from $N_{\mathrm{CX}}^{\mathrm{MC}}$, $N_{\mathrm{BG}}^{\mathrm{MC}}$, and the corresponding predicted CX cross section $\sigma_{\mathrm{CX}}^{\mathrm{MC}}$ following Eq. (\ref{eqn:xsec_calc}). $\sigma_{\mathrm{ABS}}$ was obtained by subtracting $\sigma_{\mathrm{CX}}$ from $\sigma_{\mathrm{ABS+CX}}$ obtained in \cite{duet}.

 \begin{equation} \label{eqn:xsec_calc}
 \begin{aligned}
 \sigma_{\mathrm{CX}} &= 
 \sigma_{\mathrm{CX}}^{\mathrm{MC}}
 \times \frac{N_{\mathrm{Data}}-N_{\mathrm{BG}}^{\mathrm{MC}}}{N_{\mathrm{CX}}^{\mathrm{MC}}} \\
 &\times
 \frac{1-R_{\mathrm{TiO}_2}^{\mathrm{Data}}}{1-R_{\mathrm{TiO}_2}^{\mathrm{MC}}}
 \times \frac{1}{1-f_{\mu}},
 \end{aligned}
 \end{equation} 

Corrections for the fraction of muons in the beam ($f_{\mu}$) and the fraction of interactions on TiO$_2$ nuclei ($R_{\mathrm{TiO}_2}^{\mathrm{Data}}$ and $R_{\mathrm{TiO}_2}^{\mathrm{MC}}$) were also applied. 

\subsection{Selection, rejection and mis-reconstruction efficiencies}\label{sec:efficiencies}

\begin{enumerate}
\item{{\bf $\pi^0$ selection efficiency:} the probability of a true CX event passing the selection criteria as a function of the outgoing $\pi^0$ momentum and angle is defined as the ratio of the distributions before and after the selection is applied. This selection efficiency is shown in Fig. \ref{fig:pi0_selection} for the $p_{\pi}$ = 201.6 MeV$/c$ setting as an example.}

\begin{figure}[h]
 \includegraphics[width=86mm]{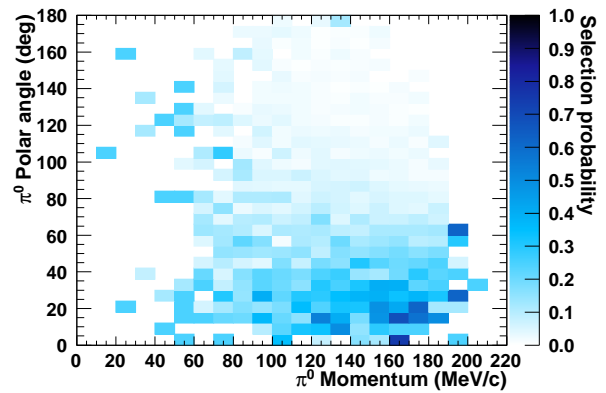}
 \caption{(Color online) Selection efficiency of true CX events as a function of the outgoing $\pi^{0}$ momentum and angle, for the $p_{\pi}$ = 201.6 MeV$/c$ setting.}
 \label{fig:pi0_selection}
\end{figure}

\item{{\bf Proton/neutron veto rejection:} the probability that an ejected proton or neutron will produce hits in the first two XY modules of CEMBALOS. Fig. \ref{fig:proton_rejection} shows the rejection efficiency for protons in the the 201.6 MeV$/c$ setting. The CEMBALOS forward acceptance ($<45^{\circ}$) can be clearly seen.}

\begin{figure}[h]
 \includegraphics[width=86mm]{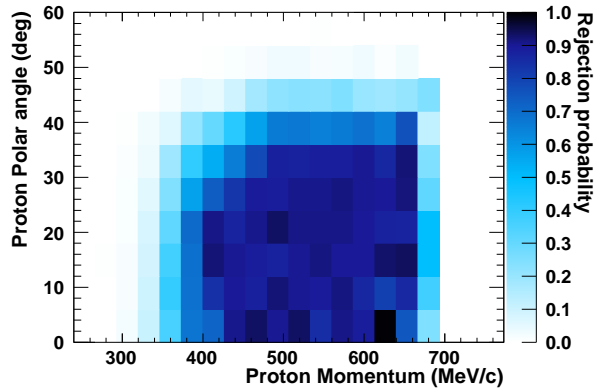}
 \caption{(Color online) Rejection probability of events where an ejected proton from ABS or quasi-elastic scattering fails the veto rejection criteria, as a function of its outgoing  momentum and angle, for the $p_{\pi}$ = 201.6 MeV$/c$ setting.}
 \label{fig:proton_rejection}
\end{figure}

\item{{\bf Proton mis-reconstruction:} the probability of a proton being mis-reconstructed as a ``pion-like'' track in PIA$\nu$O thus causing the event to be rejected.}

\item{{\bf $\pi^{+}$ mis-reconstruction and veto:} the probability of an outgoing $\pi^{+}$ following a quasi-elastic scatter to be mis-reconstructed in PIA$\nu$O as a ``proton-like'' track and then producing hits in the first two XY modules of CEMBALOS.}

\item{{\bf Neutron selection efficiency:} the probability of a neutron from an ABS event passing the selection criteria.}
\end{enumerate}

In this scheme a true CX event would be categorized as a signal event if: the $\pi^{0}$ is selected, the ejected proton(s) is not mis-reconstructed as a ``pion-like'' track in PIA$\nu$O, and the ejected nucleons do not trigger the veto rejection. On the other hand, an ABS or quasi-elastic scattering event would be categorized as a background event if: a neutron is selected, any outgoing $\pi^{+}$ is mis-reconstructed as a proton, all ejected proton(s) are not mis-reconstructed in PIA$\nu$O as ``pion-like'', and the the ejected nucleons do not trigger the CEMBALOS veto rejection. 

\subsection{Selection of nominal model}\label{sec:nominal}

The results of applying this scheme to model predictions from various models are summarized in Table \ref{tbl:eff_scheme_results} for each momentum setting. In addition to \textsc{Neut} and \textsc{Fluka}, the scheme was applied to the \textsc{Geant4} model prediction calculated from a thin target simulation (independent of the DUET simulation) as a means of validation of the procedure. The predictions of $N_{\mathrm{CX}}^{\mathrm{MC}}$ for \textsc{Geant4} agree with $N_{\mathrm{CX}}^{\mathrm{G4}}$ from Table \ref{tbl:short_event_summary} within $\sim$3\%, while $N_{\mathrm{BG}}^{\mathrm{MC}}$ were underestimated as not all sources of background were included in the scheme. These are discussed in Sec. \ref{sec:background}.
\begin{table}[htbp]
\begin{center}
\begin{tabular}{c|c|c|c|c|c}
\hline
$p_{\pi}$ [MeV$/c$] & Model & $\sigma_{CX}^{MC}$ [mb] &  $N_{\mathrm{CX}}^{\mathrm{MC}}$  &  $N_{\mathrm{BG}}^{\mathrm{MC}}$  &  $\sigma_{CX}$ [mb] \\ \hline
\multirow{4}{*}{201.6} 
& \textsc{Geant4} & 36.7 & 63.3 & 6.1 & 58.0 \\
& \textsc{Fluka} & 55.5 & 122.2 & 6.3 & 45.3 \\
& \textsc{Neut} & 50.5 & 83.0 & 4.5 & 61.8 \\ \hline

\multirow{4}{*}{216.6} 
& \textsc{Geant4} & 37.5 & 16.5 & 2.0 & 41.6 \\
& \textsc{Fluka} & 59.5 & 32.5 & 1.5 & 34.4  \\
& \textsc{Neut} & 55.7 & 24.2 & 1.5 & 43.5 \\ \hline

\multirow{4}{*}{237.2} 
& \textsc{Geant4} & 39.6 & 80.0 & 9.7 & 65.4 \\
& \textsc{Fluka} & 61.7 & 149.4 & 5.8 & 56.1 \\
& \textsc{Neut} & 57.5 & 111.7 & 6.1 & 69.8 \\ \hline

\multirow{4}{*}{265.5} 
& \textsc{Geant4} & 44.7 & 88.8 & 9.6 & 71.4 \\
& \textsc{Fluka} & 62.4 & 143.5 & 5.0 & 63.7 \\
& \textsc{Neut} & 57.9 & 129.4 & 6.9 & 64.8 \\ \hline

\multirow{4}{*}{295.1} 
& \textsc{Geant4} & 45.1 & 122.5 & 12.7 & 55.1 \\
& \textsc{Fluka} & 58.5 & 176.2 & 5.6 & 52.0 \\
& \textsc{Neut} & 58.3 & 170.3 & 8.4 & 52.7 \\ \hline
\end{tabular}
\caption{Predicted $N_{\mathrm{CX}}^{\mathrm{MC}}$, $N_{\mathrm{BG}}^{\mathrm{MC}}$ and extracted CX cross section $\sigma_{\mathrm{CX}}$ obtained from applying the efficiency scheme to \textsc{Geant4}, \textsc{Fluka}, and \textsc{Neut} model predictions. See text for discussion.}
\label{tbl:eff_scheme_results}
\end{center}
\end{table}

The differences in the extracted cross section among models range from 21.9\% at $p_{\pi}$ = 201.6 MeV$/c$ to 5.7\% at $p_{\pi}$ = 295.1 MeV$/c$, with \textsc{Fluka} and \textsc{Geant4} being the extreme case scenarios. This is consistent with the model comparison from Fig. \ref{fig:pi0kinem}. Considering the good agreement between \textsc{Fluka} and the external data in Fig. \ref{fig:pi0kinem}, the results from applying the efficiency scheme to \textsc{Fluka}, with the $N_{\mathrm{BG}}^{\mathrm{MC}}$ prediction scaled up to increase the additional backgrounds not included in the scheme, were chosen as our nominal result.

%% file: uncertainties.tex
\section{\label{sec:uncertainties}Systematic Uncertainties}
Multiple sources of systematic errors were investigated. Estimation procedures for beam and PIA$\nu$O detector systematics are unchanged from \cite{duet} and are briefly outlined in Sec. \ref{sec:beam_syst} and \ref{sec:piano_syst}. CEMBALOS detector systematics are summarized in Sec. \ref{sec:cembalos_syst}. Uncertainties related to the physics modeling are discussed in Sec. \ref{sec:physics_syst}. Table \ref{table:systematics} shows a summary of all the systematic uncertainties estimated for this analysis.

\begin{table*}[htbp]
\begin{center}
\begin{tabular*}{\textwidth}{l|@{\extracolsep{\fill}}ccccc|ccccc}
\hline\hline
& \multicolumn{5}{c}{CX} & \multicolumn{5}{c}{ABS} \\
\hline
{\bfseries$\pi^+$ Momentum [MeV/c]}& 201.6 & 216.6 & 237.2 & 265.5 & 295.1 & 201.6 & 216.6 & 237.2 & 265.5 & 295.1 \\
\hline
  {\bfseries Beam Systematics} & & & & &  & & & & &\\
  ~~Beam profile& 3.5& 4.9& 6.2& 4.2& 2.0& 2.2& 2.7& 3.8& 2.9& 2.5 \\
  ~~Beam momentum& 4.1& 1.6& 3.5& 4.1& 2.8& 1.5& 2.3& 1.9& 2.5& 3.0 \\
  ~~Muon Contamination& 0.5& 0.8& 0.9& 0.3& 0.2& 0.5& 0.8& 0.9& 0.3& 0.2 \\
  \hline
  {\bfseries PIA$\nu$O Systematics} & & & & &  & & & & &\\
  ~~Fiducial volume& 3.6& 2.3& 4.3& 3.9& 4.5& 1.1& 5.4& 4.1& 3.8& 3.4 \\
  ~~Charge distribution& 3.3& 4.1& 3.3& 2.4& 3.0& 4.3& 3.2& 4.1& 4.1& 4.4 \\
  ~~Crosstalk probability& 3.9& 4.9& 4.4& 2.5& 2.2& 1.9& 2.0& 2.7& 1.7& 1.3 \\
  ~~Layer alignment& 1.3& 3.6& 2.9& 0.9& 1.1& 1.0& 2.3& 2.8& 1.7& 2.4 \\
  ~~Hit inefficiency& 1.0& 2.1& 2.1& 2.5& 2.6& 1.1& 1.3& 1.5& 2.0& 1.0 \\
  ~~Target material& 2.0& 2.0& 2.9& 2.9& 2.9& 1.2& 1.2& 1.2& 1.2& 1.3 \\
  \hline
  {\bfseries CEMBALOS Systematics} & & & & &  & & & & &\\
  ~~Charge calibration& 1.7& 1.6& 3.7& 3.1& 6.7& 1.3& 1.1& 2.0& 1.7& 2.5 \\
  ~~Hit inefficiency& 1.6& 2.1& 1.1& 1.3& 2.0& 1.2& 1.1& 1.1& 1.0& 0.9 \\
  ~~Position and alignment & 7.7& 7.9& 8.3& 5.7& 4.6& 0.7& 1.0& 0.7& 0.7& 1.0 \\
  \hline
  {\bfseries Physics Systematics} & & & & &  & & & & &\\
  ~~$\pi^{0}$ kinematics& 6.1& 6.9& 7.9& 9.4& 10.6& 2.1& 1.6& 3.2& 4.3& 4.1 \\
  ~~Nuclear de-excitation $\gamma$ background& 0.9& 0.8& 0.7& 0.6& 0.6& 0.4& 0.2& 0.7 & 0.3& 0.2 \\
  ~~Multiple interactions& 1.1 & 1.9 & 1.7 & 1.5 & 1.8 & 0.5& 0.5& 0.8& 0.7& 0.7 \\
  ~~Pion decay background& 1.9 & 2.8& 1.2& 0.6& 0.9& 0.8& 0.7& 0.5& 0.3& 0.3 \\
  \hline
  {\bfseries Statistical error} & 11.0& 26.0& 9.4& 8.9& 8.8& 3.9& 6.2& 3.9& 4.2& 3.6 \\
  \hline\hline
  {\bfseries ~~Total error} & 17.9& 30.3& 19.4& 17.0& 18.0& 7.8& 10.5& 10.4& 9.7 & 9.6 \\
  \hline
\end{tabular*}
\caption{Summary of the statistical and systematic uncertainties in percentage.}
\label{table:systematics}
\end{center}
\end{table*}

\subsection{Beam systematics}\label{sec:beam_syst}
The pion beam profile and momentum were measured using PIA$\nu$O through-going pion data. The uncertainties were less than $\sim$1 mm and $\sim$1 MeV$/c$, respectively. The systematic error was evaluated by changing the momentum, the center position, and the beam spread in the MC within their uncertainty and re-calculating the cross sections.
\subsection{PIA$\nu$O detector systematics}\label{sec:piano_syst}
Various sources of systematic uncertainty were estimated for PIA$\nu$O following the procedures described in \cite{duet}. These account for uncertainties on the scintillator fiber composition, the size of the fiducial volume, the alignment of the fibers, and the simulation of the charge deposition, hit detection efficiency, and crosstalk. For this analysis the same procedures were used.
\subsection{CEMBALOS detector systematics}\label{sec:cembalos_syst}
\subsubsection{\bf Position and alignment}
The overall uncertainty in the position of CEMBALOS relative to PIA$\nu$O, and of the position of the scintillator and lead modules relative to the dark box as well as each other is estimated to be $\pm$5 mm. This corresponds to a change of $\sim$3.4\% in the subtended solid angle. The effect to the calculated cross section is estimated by shifting the position of CEMBALOS in the simulation $\pm$ 5mm in the $x,y$, and $z$ directions. The relatively large size of this systematic uncertainty (4.5$\sim$8.3\%) is due to the sensitivity of this measurement to the $\pi^0$ kinematics and will be discussed in further detail in Sec. \ref{sec:physics_syst}.
\subsubsection{\bf Charge simulation}
\begin{figure}[ht]
 \includegraphics[width=86mm]{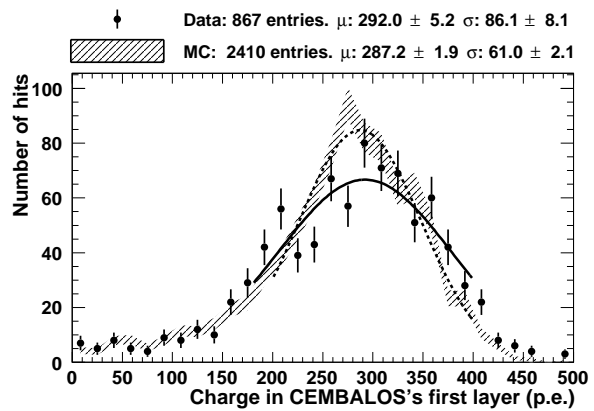}
 \caption{Charge distribution in the first layer of CEMBALOS for stopping protons in the 237.2 MeV$/c$ setting for data (circles) and MC (filled histogram). Only statistical uncertainties are plotted. The solid and dashed lines are Gaussian fits to data and MC respectively.}
 \label{fig:proton_sample}
\end{figure}

The calibration procedure presented in Sec. \ref{section:calibration} and Fig. \ref{fig:muoncharge} show that for single hits from minimum ionizing particles ($<50$ p.e.) the charge simulation agrees with data at the $\sim5$\% level. However, as can be seen from Fig. \ref{fig:nhitsvsCharge}, for most of the background events the charge deposited per hit is above this region. A control sample of protons stopping in the first two XY modules of CEMBALOS was used to estimate the accuracy of the charge simulation for higher energy depositions. It was obtained by using $dQ/dx$ information from PIA$\nu$O to select ``proton-like'' tracks and requiring all CEMBALOS hits to be in the first two XY modules. Fig. \ref{fig:proton_sample} shows the charge deposition distribution in the first layer of CEMBALOS for this sample in data and MC.

The discrepancy in the resolution for data and MC was estimated to be 20\% from the widths of Gaussian fits of the distributions in Fig. \ref{fig:proton_sample}.
A random Gaussian smearing with a 20\% width was applied to the charge deposited by each hit in every event for 1000 toy MC experiments to determine what fraction of the time signal events were mis-reconstructed as background and vice versa. The cross section was calculated for each toy experiment and the spread was taken as the uncertainty.

\subsubsection{\bf Hit inefficiency}
\begin{figure}[ht]
 \includegraphics[width=86mm]{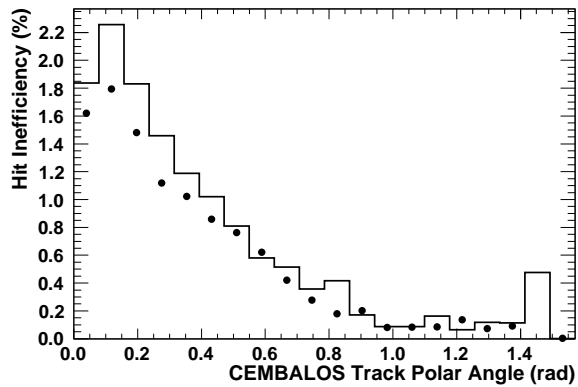}
 \caption{CEMBALOS hit inefficiency for data (circles) and MC (solid line) in the $p_\pi=$237.2 MeV$/c$ setting. The statistical error bars are too small to appear.}
 \label{fig:hit_ineff}
\end{figure}

The hit reconstruction inefficiency in CEMBALOS was measured by counting how often a hit was missing in a reconstructed track. The tracks were required to have at least two hits in both the first and last two layers. Fig. \ref{fig:hit_ineff} shows the hit inefficiency, defined as the ratio of missing hits over the total number of hits expected, for data and MC in the 237.2 MeV$/c$ setting as a function of the CEMBALOS reconstructed polar angle. The hit inefficiency integrated over all angles is 1.33\% and 1.16\% for data and MC respectively.

The effect on the measured cross section is estimated by randomly deleting CEMBALOS hits in 1000 MC toy experiments with a probability given by the difference of the integrated hit inefficiencies for data and MC, affecting both the hit multiplicity and total charge deposited.

\subsection{Physics modeling systematics}\label{sec:physics_syst}
\subsubsection{\bf{ Uncertainty from $\pi^{0}$ kinematics }}
True CX events were reweighted following the discrepancy between \cite{Ashery2} and the \textsc{Fluka} model prediction as a function of the $\pi^{0}$ angle. The weights ranged from 0.7 to 1.3, while the average weight applied was 0.9. The effect on $\sigma_{\mathrm{CX}}$ ranged  from 6.1\% to 10.6\%, representing the largest systematic uncertainty for this analysis.

\subsubsection{\bf Other backgrounds}\label{sec:background}
The uncertainties from additional contributions to the number of predicted background events were estimated in three different categories, as described in the following text.

{ \it Nuclear de-excitation $\gamma$-rays:} inelastic interactions can leave the nucleus in an excited state. Low-energy ($<25$ MeV$/c$) $\gamma$-rays can be emitted as the nucleus returns to its ground state. If these photons interact in CEMBALOS they can fake a signal event. While these processes are believed to be well modeled by our simulation, we assign a conservative 100\% error on the number of background events from this process.\\

{ \it Multiple interactions: } it is possible for the initial $\pi^{+}$ to be scattered (both elastically or quasi-elastically) before it undergoes a CX interaction. The CX interaction can take place inside the PIA$\nu$O FV ($\sim$58\%), outside the FV but still in a scintillator fiber ($\sim$37\%), or somewhere in the aluminum support structure and/or dark boxes of PIA$\nu$O or CEMBALOS ($\sim$5\%). The uncertainty of the number of events of this type of background event is estimated from the uncertainty on elastic and CX interactions on carbon and aluminum from previous experiments.\\

{ \it $\pi^{+}$ decay products: } a $\pi^{+}$ that scatters in PIA$\nu$O and produces a fake ``proton-like'' track can then stop and decay around or inside CEMBALOS, possibly circumventing the veto rejection. The decay products can then deposit enough energy in CEMBALOS to produce a fake signal event. A conservative 100\% uncertainty is assigned to these events, which amount to $\sim$1\% of the selected events. 



%% file: result.tex
\section{Results}\label{sec:result}
\begin{figure}[h]
\begin{center}
\includegraphics[width=86mm]{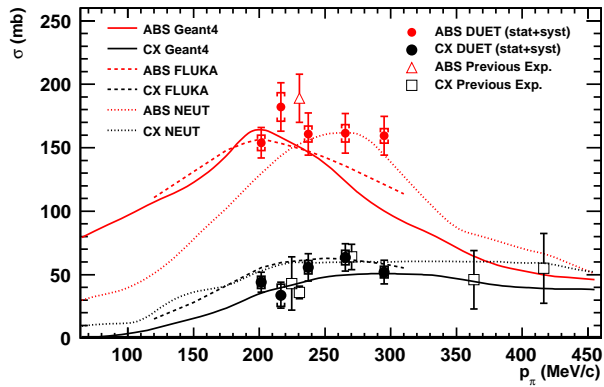}
\caption{(Color online) DUET measurements of $\sigma_{\mathrm{ABS}}$ and $\sigma_{\mathrm{CX}}$ compared with previous measurements \cite{Bellotti1973,Ashery2,Bellotti1973_2,Jones} and ABS (red) and CX (black) model predictions from \textsc{Geant4} (solid line), \textsc{Fluka} (dashed line) and \textsc{Neut} (dotted line).}
\label{fig:result}
\end{center} 
\end{figure}

The measured $\sigma_{\mathrm{ABS}}$ and $\sigma_{\mathrm{CX}}$ are presented in Table \ref{tbl:result} and shown in Fig. \ref{fig:result} with statistical and systematic error bars as a function of pion momentum, compared with the results from previous experiments \cite{Bellotti1973,Ashery2,Bellotti1973_2,Jones}. Our results are in agreement with previous experiments, but we have extended the momentum region over which the data is presented. As summarized in Table \ref{table:systematics}, the total error is $\sim$9.5\% for $\sigma_{\mathrm{ABS}}$ and $\sim$18\% for $\sigma_{\mathrm{CX}}$, except for the $p_{\pi}$ = 216.6 MeV$/c$ data set.

\begin{table}[h]
   \begin{tabular}{c|c|c}
    \noalign{\hrule height 1pt}
    $p_{\pi}$  [MeV$/c$] & $\sigma_{\mathrm{ABS}}$ [mb] & $\sigma_{\mathrm{CX}}$ [mb]\\\hline
    201.6 & 153.8 $\pm$ 12.0 & 44.0 $\pm$ 7.9 \\
    216.6 & 182.1 $\pm$ 19.2 & 33.8 $\pm$ 10.2 \\
    237.2 & 160.8 $\pm$ 16.6 & 55.8 $\pm$ 10.8 \\
    265.6 & 161.4 $\pm$ 15.7 & 63.5 $\pm$ 10.8 \\
    295.1 & 159.4 $\pm$ 15.3 & 52.0 $\pm$ 9.3\\
    \noalign{\hrule height 1pt}
   \end{tabular}
\caption{$\sigma_{\mathrm{ABS}}$ and $\sigma_{\mathrm{CX}}$ measured by DUET.}
\label{tbl:result}
\end{table}

\subsection{Fractional covariance and correlation coefficients}
\begin{figure}[h]
\begin{center}
\includegraphics[width=86mm]{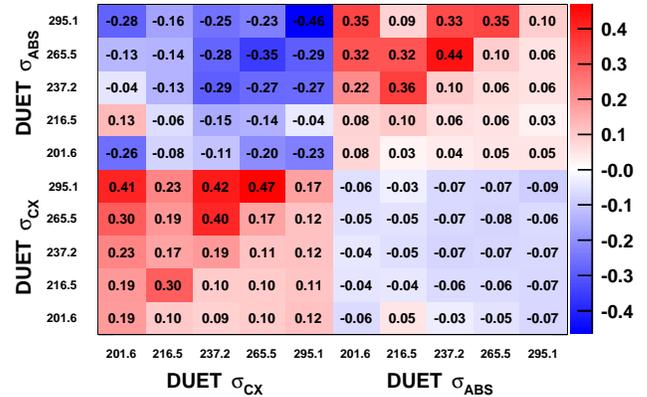}
\caption{(Color online) Fractional covariance and correlation for the DUET measurements of $\sigma_{\mathrm{ABS}}$ and $\sigma_{\mathrm{CX}}$. The  diagonal and lower triangle show the covariance $(\mathrm{Sign}(V_{ij})*\sqrt{V_{ij}})$, while the upper triangle of the matrix shows the correlation coefficients.}
\label{fig:covariance}
\end{center} 
\end{figure}

We provide the fractional covariance and correlation coefficients for the 5 $\sigma_{\mathrm{ABS}}$ and 5 $\sigma_{\mathrm{CX}}$ measured data points in the matrix in Fig. \ref{fig:covariance}. The diagonal and lower triangle of the matrix show the fractional covariance $(\mathrm{Sign}(V_{ij})*\sqrt{V_{ij}})$, where $V_{ij} = (\Delta\sigma_{i}\Delta\sigma_{j})/(\sigma_{i}\sigma_{j})$, and $\sigma_{k}$ and $\Delta\sigma_{k}$ are the nominal cross sections and their systematic shift, respectively. The upper triangle of the matrix shows the correlation coefficients.
The statistical uncertainties were included as an uncorrelated diagonal matrix. There are positive correlations within the $\sigma_{\mathrm{ABS}}$ and $\sigma_{\mathrm{CX}}$ measurements, and negative correlations across them, as is expected from the subtraction method used. This is the first time that a correlation matrix is published for a pion inelastic cross section measurement. 

\section{Summary}
We obtained $\sigma_{\mathrm{ABS}}$ and $\sigma_{\mathrm{CX}}$ for positive pions in carbon nuclei at five incident momenta between 201.6 MeV$/c$ to 295.1 MeV$/c$. A covariance matrix for the 10 measured data points was produced. This result will be an important input to existing models such as \textsc{Geant4} or \textsc{NEUT} to constrain sub-GeV pion interactions.